\documentclass[journal]{IEEEtran}
\ifCLASSINFOpdf
\else
\fi

\usepackage[utf8]{inputenc}
\usepackage[pdftex]{graphicx}
\usepackage{url}
\usepackage{xspace}
\usepackage{cite}
\usepackage{amsmath}
\usepackage{amssymb}
\usepackage{balance}
\usepackage{tabularx}
\usepackage{booktabs}
\usepackage{siunitx}
\usepackage{float}
\usepackage{subcaption}
\newcommand{\et}{\emph{et al.}\xspace}
\def\code#1{\texttt{#1}}

\newlength{\bibitemsep}
\newlength{\bibparskip}
\let\oldthebibliography\thebibliography
\renewcommand\thebibliography[1]{%
  \oldthebibliography{#1}%
  \setlength{\parskip}{\bibitemsep}%
  \setlength{\itemsep}{\bibparskip}%
}
\begin{document}
%
\title{Tiny Transformers for Environmental Sound Classification at the Edge}
%
%
%

\author{David~Elliott,~\IEEEmembership{Member,~IEEE,}
        Carlos~E.~Otero,~\IEEEmembership{Senior~Member,~IEEE,}
        Steven~Wyatt,~\IEEEmembership{Member,~IEEE,}
        Evan~Martino,~\IEEEmembership{Member,~IEEE}
\thanks{D. Elliott, C. E. Otero, S. Wyatt, and E. Martino are with the Department
of Computer Engineering and Sciences, Florida Institute of Technology, Melbourne,
FL, 32901 USA e-mail: delliott2013@my.fit.edu.}

\thanks{This work has been submitted to the IEEE for possible publication. Copyright may be transferred without notice, after which this version may no longer be accessible.}%
}

\maketitle

\begin{abstract}
With the growth of the Internet of Things and the rise of Big Data, data processing and machine learning applications are being moved to cheap and low size, weight, and power (SWaP) devices at the edge, often in the form of mobile phones, embedded systems, or microcontrollers. The field of Cyber-Physical Measurements and Signature Intelligence (MASINT) makes use of these devices to analyze and exploit data in ways not otherwise possible, which results in increased data quality, increased security, and decreased bandwidth. However, methods to train and deploy models at the edge are limited, and models with sufficient accuracy are often too large for the edge device. Therefore, there is a clear need for techniques to create efficient AI/ML at the edge. This work presents training techniques for audio models in the field of environmental sound classification at the edge. Specifically, we design and train Transformers to classify office sounds in audio clips. Results show that a BERT-based Transformer, trained on Mel spectrograms, can outperform a CNN using 99.85\% fewer parameters. To achieve this result, we first tested several audio feature extraction techniques designed for Transformers, using ESC-50 for evaluation, along with various augmentations. Our final model outperforms the state-of-the-art MFCC-based CNN on the office sounds dataset, using just over 6,000 parameters -- small enough to run on a microcontroller.
\end{abstract}


%
\IEEEpeerreviewmaketitle

\section{Introduction}
%
%
%
%

\IEEEPARstart{T}{he} field of environmental sound classification (ESC) has been actively researched for many years, with applications in security, surveillance, manufacturing, AVs, and more \cite{cowling2003comparison}. In modern days, ESC has important applications to autonomous vehicles (AV), as they can be used to detect sirens, accidents, locations, in-cabin disturbances, and much more. As vehicle-based computational power increases, and algorithms improve, it becomes vital to explore a wide number of options to perform a given machine learning task. For ESC, this means exploring transformers \cite{vaswani2017attention} as a possible means to perform ESC at the edge.

Recent work in transformers has profoundly affected the field of natural language processing (NLP), seeing models such as BERT \cite{devlin2018bert}, XLNet \cite{yang2019xlnet}, T5 \cite{raffel2019exploring}, GPT \cite{radford2018improving,radford2019language,brown2020language}, and BigBird \cite{zaheer2020big} -- to name a few -- iteratively setting new state-of-the-art for a variety of difficult NLP tasks. In many cases, transformer accuracy exceeds the performance of humans on the same tasks. 

Even though most of the highly public work with transformers has been done in NLP, a transformer, which fundamentally is simply a series of self-attention operations stacked on top of one another \cite{vaswani2017attention}, is a general architecture that can be applied to any input. OpenAI, an AI research and development company focused on ensuring artificial general intelligence benefits humanity \footnote{\url{https://openai.com/}}, made this clear in several of their recent works. In ImageGPT \cite{chen2020generative}, Chen \et showed how the GPT architecture, which is transformer-based, can be trained in an autoregressive fashion on a wide variety of images, in order to generate realistic image completions and samples. Notably, images were restricted to 64x64 pixels, as a greater amount of pixels requires substantially more compute than was feasible. In Jukebox \cite{dhariwal2020jukebox}, a transformer is used along with vector-quantized variational autoencoders (VQ-VAEs) \cite{van2017neural}, in order to generate realistic audio. Also notable is the fact that Dhariwal \et trained the transformer on a highly compressed representation of the audio waveform generated by VQ-VAEs, rather than the raw waveform, and that the outputs from the transformer are not used directly, but are passed through an upsampler first. Even so, the total cost of training the full Jukebox system is in excess of 20,000 V100-days -- an enormous cost \cite{dhariwal2020jukebox}. More recently, in a paper under review at ICLR 2021, it has been found that, given enough data (hundreds of millions of examples), transformers can exceed even the best CNNs in accuracy \cite{anonymous2021an}. We take this, in addition to the recent trend in larger datasets and more compute, to mean that any work we perform here with small datasets and modest transformers can easily be scaled up at a later date.

The field of audio speech recognition (ASR) has picked up transformers with vigor. Indeed, it is understandable, as applying transformers is straightforward for many approaches to ASR. Often based on encoder-decoders \cite{pham_very_2019}, the most obvious use of a transformer is in the decoder of an ASR system, which usually has text as input and output, thus making the application of any state-of-the-art language models, such as BERT or its variants, available to it with little adaptation needed. Some work \cite{zhang_transformer_2020} has also made use of a transformer in the encoder as well, thus making the ASR system an end-to-end transformer model. Recent work has seen transformers improve the state-of-the-art for ASR by reducing word error rate by 10\% in clean speech, and 5\% in more challenging speech\cite{shi_weak-attention_2020}.


ESC, in some ways, is much simpler than ASR, as it is not concerned about both text and audio, nor does it need to perform fine-grained classification of words or sounds per variable segment of time. However, in other ways, there are more challenges with ESC than ASR. The first major challenge that ESC presents is one of data availability; there is very little data for ESC tasks available, and even the largest and most accurate (DCASE\footnote{\url{http://dcase.community/}}\footnote{\url{https://www.kaggle.com/c/dcase2018-task1a-leaderboard}}) is only tens of thousands of audio files \cite{stowell_dcase_2015}. In addition, there is no agreed-upon standard for what sounds make up ESC. Different ESC datasets often have overlap, and Google's AudioSet \cite{gemmeke2017audio} provides the largest set of audio labels to date, but many of AudioSet's labels have to do with music or speech, and are not necessarily ``environmental''. Additionally, ESC datasets are highly heterogeneous, having an extremely broad range of sounds, varying in length, frequency, and intensity. This can increase the difficulty that a machine learning model has in learning the sounds, as they may vary widely from clip to clip.


Environmental sound classification has recently been performed mostly by convolutional neural networks (CNNs) \cite{dai_very_2017,salamon_deep_2017,tokozume_learning_2017,zhang_dilated_2017,abdoli_end--end_2019,khamparia_sound_2019,sharma_environment_2020}. These networks vary somewhat in structure, with some being fully convolutional and able to take varying-length input \cite{kumar_knowledge_2018}, others making use of modified ``attention'' layers to boost performance \cite{zhang_learning_2019}, and others using deep CNNs \cite{khamparia_sound_2019,sharma_environment_2020}. Performance on ESC is typically measured using ESC-50 \cite{piczak2015esc}, ESC-10 \cite{piczak2015esc}, Urban-8k \cite{salamon2014dataset}, or DCASE 2016-2018 \cite{giannoulis_2013}. Top reported performance on ESC-50, to the best of our knowledge, is 88.50\%, by Sharma \et \cite{sharma_environment_2020}, using a spatial attention module and various augmentations. Human accuracy has been measured at 81.3\% \cite{piczak2015esc}. We use ESC-50 in this work to allow comparison of our different models in the first stage of our research, in Section \ref{sec:experiments-esc50}.



Unfortunately, state-of-the-art transformers are too large to run on edge devices, such as mobile phones, as they often contain billions of parameters. The largest model of GPT-3 \cite{brown_language_2020} contains 175 billion parameters. When most mobile phones contain several GBs of RAM, any model exceeding one billion parameters (which, when quantized, is 1 billion bytes, or 1 GB) is likely to be inaccessible. When considering that many microcontrollers have only kilobytes of SRAM, it becomes particularly obvious that state-of-the-art transformers are not yet edge-ready. There are also latency and power considerations, as the sheer number of computations by such large models pose a limitation on how quickly a forward pass can be computed using the available processors on the device, and may drain the device's battery very quickly, or exceed power requirements. There are methods to perform knowledge distillation on transformer models, to produce models with slightly reduced accuracy, but substantially smaller in size \cite{sanh_distilbert_2020}. However, these methods require the existence of a pretrained model alongside or from which they can learn \cite{sun_patient_2019}, which does not exist in the field of ESC. 


In this work, we attempt to tackle some of these problems. For simplicity, to lower costs, and to improve the iteration time of our development process, we restrict our models to a modest size for most of our analyses. This approach, though motivated by limited resources, is supported by literature, as larger models can always be built later, after promising avenues have been discovered \cite{turc2019well}.

Our contributions in this work are as follows:

\begin{enumerate}
    \item We provide a thorough evaluation of transformer performance on ESC-50 using various audio feature extraction methods.
    \item For the most promising feature extraction method, we perform a Bayesian search through the hyperparameter space to find an optimal configuration.
    \item Based on the optimal model discovered through the Bayesian search, we train transformers on the Office Sounds dataset, and obtain a new single-model state of the art. We also train a 6,000-parameter model that exceeds the accuracy of a much larger MFCC-based CNN. 
    \item We test selected models' performance on a mobile phone, and report results.
\end{enumerate}

\begin{figure*}[t]
  \includegraphics[width=1.0\textwidth, keepaspectratio]{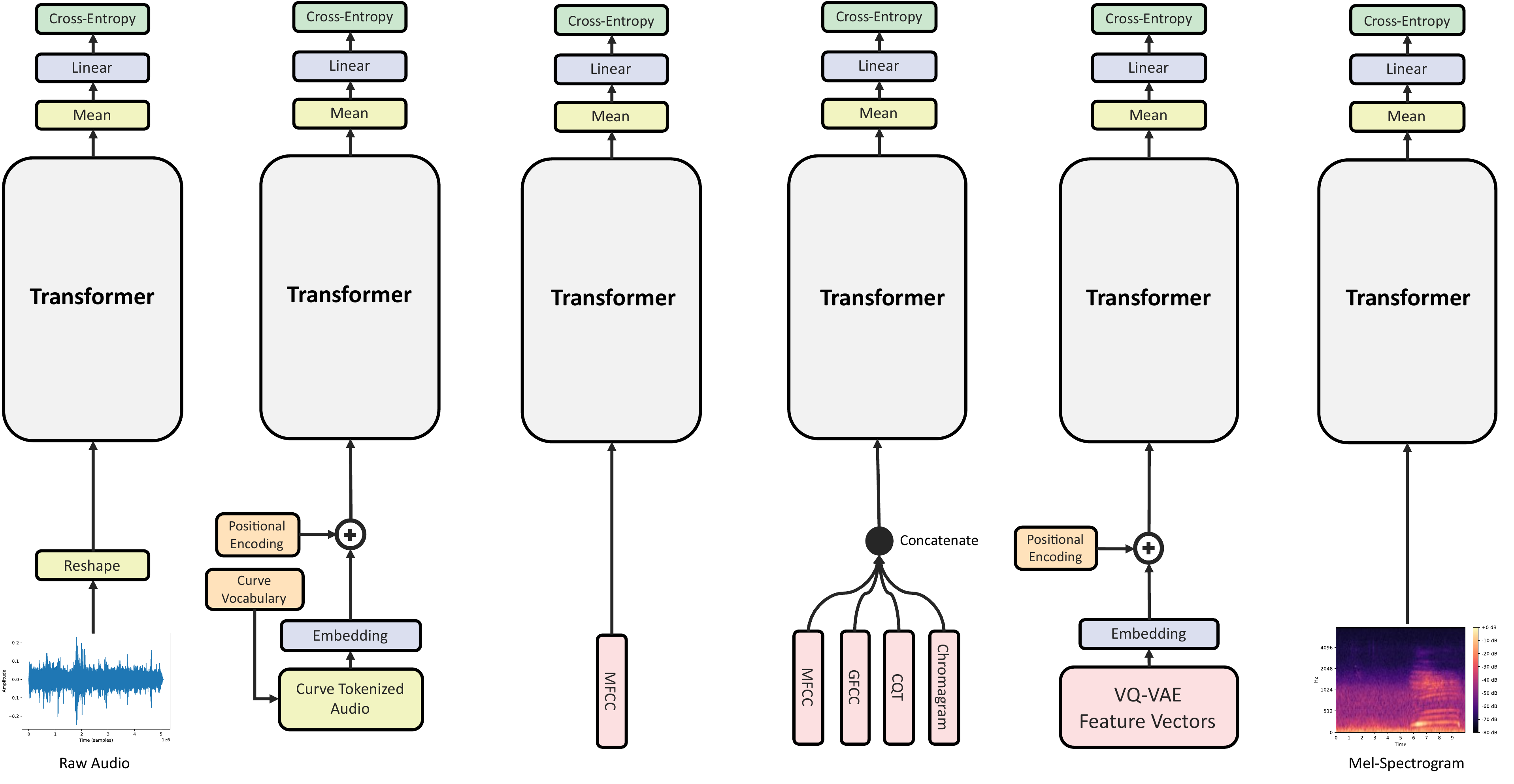}
  \caption{We take six unique approaches to training a transformer on ESC. From left to right: raw amplitude reshaping, curve tokenization, MFCC feature extraction, multi-feature extraction, VQ-VAE tokenization, and Mel spectrograms. The transformer architecture is common between them, but the first layers vary. If a positional encoding is not shown for an architecture, then no positional encoding is used.}
  \label{fig:approaches}
\end{figure*}

\section{Related Work}


There have been many attempts to classify environmental sounds accurately, At first many of the attempts used a more algorithmic approach \cite{couvreur_automatic_nodate}, focusing on hand-crafted features and mechanisms to process sound and produce a classification. However, CNNs, which had been shown to perform well on image recognition tasks, eventually became the state of the art on ESC. The current reported state of the art on the ESC-50 dataset is by Sharma \et who used a deep CNN with multiple feature channels (MFCC's, GFCC's, CQT's and Chromagram) and data augmentations as the input to their model \cite{sharma_environment_2020}. They achieved a score of 97.52\% on UrbanSound8K, 95.75\% on ESC-10, and 88.50\% on ESC-50. (Note that the original publication of Sharma \et's work included a bug in their code, which resulted in a much higher reported accuracy. That has since been corrected, but has been cited incorrectly at least once \cite{mushtaq172spectral}.) Additionally, a scoreboard has been kept in a GitHub repository \footnote{\url{https://github.com/karolpiczak/ESC-50}}, but appears to be out of date. 

Very little work has been performed with transformers on ESC tasks. Dhariwal et al. \cite{dhariwal_jukebox_2020} used transformers in an auto-regressive manner to generate music (including voices) by training on raw audio. Miyazaki et al. \cite{miyazaki_weakly-supervised_2020} proposed using transformers for sound event detection in a weakly-supervised setting. They found that this approach outperformed the CNN baseline on the DCASE2019 Task4 dataset, but no direct application of transformers to ESC has been found.
In audio, there has been a wide number of feature extraction methods used. Mitrovic \et \cite{mitrovic2010features} organized the types of feature extraction into six broad categories: temporal domain, frequency domain, cepstral domain, modulation frequency domain, phase space, and eigen domain. Features that have been used to date include Mel Frequency Cepstral Coefficients (MFCCs) \cite{boddapati_classifying_2017}, log Mel-spectrogram \cite{piczak_environmental_2015}, pitch, energy, zero-crossing rate, and mean-crossing rate \cite{li_comparison_2017}. Tak \et \cite{shankar_novel_2017} achieved state of the art accuracy with phase encoded filterbank energies. Agrawal \et \cite{agrawal_novel_2017} determined that Teager Energy Operator-based Gammatone features outperform Mel filterbank energies. To combat noisy signals, Mogi and Kasai \cite{mogi_noise-robust_2012} proposed the use of Independent Component Analysis and Matching Pursuit, a method to extract features in the time domain, rather than the frequency domain. With reference to \cite{mogi_noise-robust_2012}, we note the assumption in our work that noise in an office environment will be minimal. Sharma \et \cite{sharma_environment_2020} obtain state of the art using MFCC, GFCC, CQT, and Chromagram features. Jukebox \cite{dhariwal2020jukebox} was trained to differentiate music and artist styles using features extracted from three different VQ-VAEs, with varying vector lengths for each. A survey was performed in 2014 by Chachada and Kuo \cite{chachada_environmental_2014} that enumerated the features used in literature, with comparisons between each, but no more recent survey has been found. We choose some of the most successful of those feature extraction methods, and attempt some new ones designed specifically for transformers.

\section{Models}
\label{sec:models}


\begin{figure}[ht]
  \centering
  \includegraphics[width=0.23\textwidth, keepaspectratio]{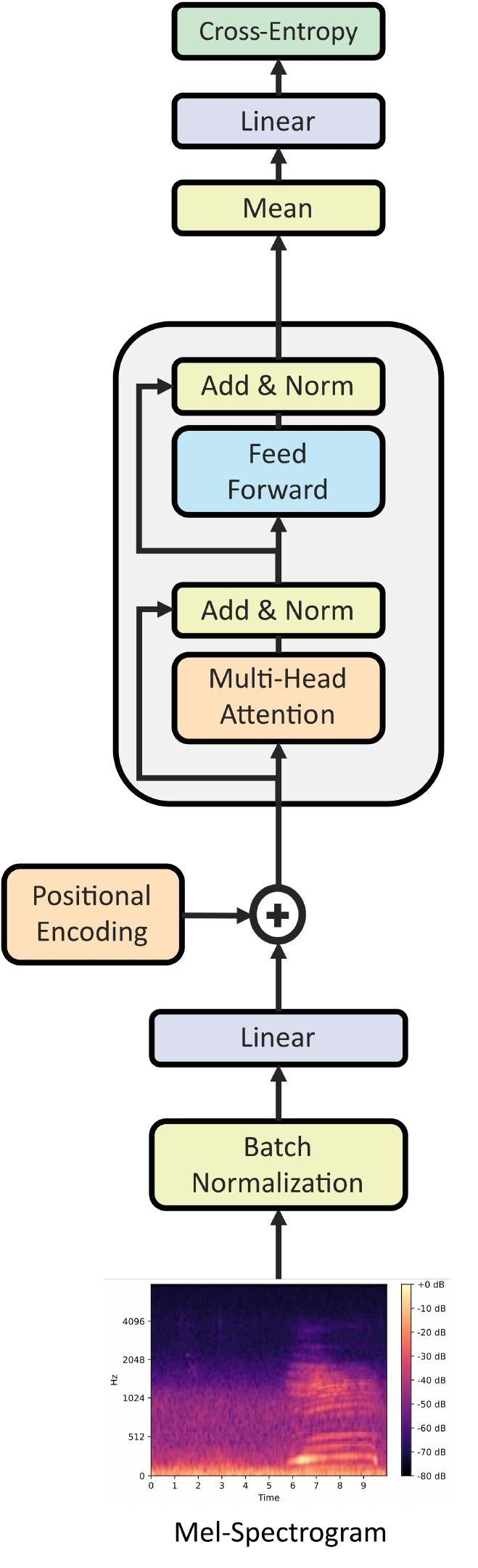}
  \caption{The architecture on which our smallest Transformer model is based. Visualization based on the diagram by Vaswani \et \cite{vaswani2017attention}.}
  \label{fig:tiny-transformer}
\end{figure}

Transformers are neural networks based on the self-attention mechanism, which is an operation on sequences, relating positions within the sequence in order to compute its representation \cite{vaswani2017attention}. We emphasize that the attention mechanism operates on a list of sequences, which means that the input to a transformer must be 2-dimensional (excluding batches). In NLP, we want the transformer to operate on sequences of words, characters, or something similar, which we refer to as ``tokens''. Therefore, in order to meet the 2-dimensional input requirements of the transformer, each token must be converted to a sequence. This is traditionally done using an embedding layer, which takes a token, represented by the integer value of the token's position in a pre-computed vocabulary, and looks up its corresponding vector representation in a matrix. This embedding matrix is able to be learned. The embedding vector length is typically referred to, in transformers, as the ``hidden size'', or $H$. The number of tokens is referred to by the length of the input $L$, also referred to as the sequence length. In this way, $H$ defines the number of dimensions that are used to represent tokens -- where more dimensions typically mean greater learning capacity -- and $L$ determines the context length, or window size, of the input.

Audio is represented at an extremely fine-grained level of detail (many samples per second), which poses challenges that NLP does not have to face. For example, the common sampling rate of 44.1 kHz in a 5-second audio clip (the length of an audio clip in ESC-50) results in 220,500 samples. Combine this with the limitations of modern-day transformers, which, with some exceptions, are limited to roughly $H<2000$ tokens, depending on available hardware, and the task of analyzing audio data becomes quite difficult. There is hope that this will change in the near future, with the creation of linear-scaling models like BigBird \cite{zaheer2020big} proven to have the same learning capacity as BERT, and recent improvements in AI hardware by NVIDIA. But, for the sake of our discussion and analysis, we will assume that we cannot use a transformer sequence length of more than 2048. 

This results in the maximum audio window that a transformer can view -- in the na\"ive case, where a single token is a single amplitude -- to be 2048 samples, or 0.046 seconds (46 milliseconds). Since sounds in the ESC-50 dataset often last much longer than 46 milliseconds, we must therefore abandon the na\"ive approach initially. The thought exists that it is possible to downsample the audio to make the 2048 sequence length be able to view a longer length of audio, but in practice this results in substantial information loss below 16 kHz, and reduces model accuracy. We would like our work to be constrained solely by hardware and algorithmic limitations, which have a strong likelihood of improving in the near future, rather than constrained by the information content in downsampled audio clips. Therefore, we assume a sampling rate above the Nyquist rate of 40 kHz for human-detectable audio, and, specifically, use the conventional value of 44.1 kHz in all of our analyses.

All models in this work are based on BERT \cite{devlin2018bert} or AlBERT \cite{lan_albert_2020}. The Transformer base structure, whether BERT or AlBERT, does not change in this work. The only alteration performed is to remove positional encodings for some models, which is noted in Figure \ref{fig:approaches} outside of the Transformer base. We note that our base structure in our ESC-50 experiments does not make use of an embedding layer for input tokens, as is customary in language models, and any tokenizations and embeddings that do occur are explicitly called out in Figure \ref{fig:approaches}.

We make a change to the transformer design in our second series of experiments on the Office Sounds dataset (Figure \ref{fig:tiny-transformer}), which allowed the size of the input to be decoupled from the size of the model. In the six models shown in Figure \ref{fig:approaches}, the input must be either reshaped or the features must be extracted in the shape required to create a transformer of the desired size. For example, using 128 Mel bands when calculating MFCCs resulted in a transformer that had a maximum hidden size of 128. We remove this dependency in our Office Sounds experiments by adding a mapping layer, as shown in Figure \ref{fig:tiny-transformer}. The mapping layer is simply a linear layer that takes input of any size and maps it to the size of the transformer. It also provides representational advantages, as this layer is able to be learned, similar to the embedding layer in traditional transformers.

Additionally, in our ESC-50 experiments, we normalize all inputs to a number between 0 and 1 as a preprocessing step, where input is not tokenized. We remove this normalization step in our Office Sounds experiments, in favor of a batch normalization layer \cite{ioffe2015batch}, which may also have provided representational advantages to the Transformer by being learnable.


\section{Approach}

We divide our approach below into sections on data, feature extraction, data augmentations, models, and model conversion. These methods work together to produce the results in Section \ref{sec:experiments}.

\subsection{Data}

We use three datasets in this work, AudioSet \cite{gemmeke2017audio}, ESC-50 \cite{piczak2015esc}, and the Office Sounds dataset \cite{elliott2020cyber}. AudioSet is a large-scale weakly labeled dataset covering 527 different sound types. The authors provide a balanced and unbalanced version of the dataset; we use the balanced dataset, with some additional balancing that we perform ourselves. Note that in order to train on the audio from this dataset, we had to download the audio from the sources used to compile the balanced dataset. This was an error-prone process, as not all sources from the original AudioSet are still available. More details on the datasets are available in Table \ref{table:data}. ESC-50 is a strongly labeled dataset containing 50 different sound types. Each sound category contains 40 sounds, making it a balanced dataset. The Office Sounds dataset is both an unbalanced and weakly labeled dataset, owing to its origins in DCASE, but nearly the same number of audio files as ESC-50, with slightly longer total length, and only 6 labels.

All audio files are converted to wave files, if they are not already formatted as such. We read from each file at a sampling rate of 44100 Hz, in mono.

\begin{table}[h]
\centering
\caption{Information on the datasets used for training.}
\begin{tabular}{llll}
\hline
\multicolumn{1}{|l|}{}                  & \multicolumn{1}{l|}{\textbf{\# of files}} & \multicolumn{1}{l|}{\textbf{\# of hours}} & \multicolumn{1}{l|}{\textbf{\# of audio types}} \\ \hline
\multicolumn{1}{|l|}{\textbf{AudioSet}} & \multicolumn{1}{l|}{37948}                & \multicolumn{1}{l|}{104.52}               & \multicolumn{1}{l|}{527}                        \\ \hline
\multicolumn{1}{|l|}{\textbf{ESC-50}}   & \multicolumn{1}{l|}{2000}                 & \multicolumn{1}{l|}{2.78}                 & \multicolumn{1}{l|}{50}                         \\ \hline
\multicolumn{1}{|l|}{\textbf{Office Sounds}}   & \multicolumn{1}{l|}{1608}                 & \multicolumn{1}{l|}{2.80}                 & \multicolumn{1}{l|}{6}                         \\ \hline
             
\end{tabular}
\label{table:data}
\end{table}

\subsection{Feature Extraction}
\label{sec:feature-extraction}

Feature extraction is a critical part of any machine learning architecture, and especially so for transformers. In fact, some of the critical work that went into making BERT such a success was the use of word pieces, rather than words or characters \cite{devlin2018bert}. In an attempt to discover a feature extraction method that can be of similar use in audio, we attempted several, some of which are well-known methods, others of which we have adapted to our particular use case. The approaches can be seen in Figure \ref{fig:approaches}.




\subsubsection{Amplitude Reshaping}
\label{sec:amplitude-reshaping}


Motivated by works such as WaveNet \cite{oord2016wavenet}, Jukebox \cite{dhariwal2020jukebox} and, in general, the move toward more ``pure'' data representations, we developed a method for the transformer to work with raw amplitudes. 

Using the notation in \cite{sperber_self-attentional_2018}, we reshape audio in the following way, where $X$ is a sequence of amplitudes $X=\{x_0, x_1, ..., x_n\}$, $l$ is the sequence length, and $d$ is the hidden dimension:

\begin{equation}
    X \in \mathbb{R}^{l * d \times 1} \rightarrow_{reshape} X \in \mathbb{R}^{l \times d}
\end{equation}

In this way, the amount of audio that we are able to process is a combination of the sequence length of the model, and the size of the hidden dimension. Under this reshaping operation, with $l=512$ and $d=512$, we are able to process data up to $262,144$ samples, or nearly 6 seconds.

\subsubsection{Curve Tokenization}

Curve tokenization is an audio tokenization method that we propose, based on WordPiece tokenization \cite{wu2016google} in NLP. The intuition behind this method is that, since audio signals typically vary smoothly, there may exist a relatively small number of ``curves'' that can describe short sequences of audio signals. These curves are commonly represented in audio signals by sequences of floating point numbers. In wave files, a single audio amplitude can be one of 65,536 values, or $2^{16}$, values; as such, our audio is, effectively, already quantized. We term the quantization level of the audio the resolution $R$. 

Although wave file audio is already quantized, it is advantageous to quantize it further, as doing so reduces the maximum number of theoretical curves that can exist within any given sequence of audio. As an example, an 8-token curve with $R=100$ has a maximum number of theoretical curves of $100^8$. We performed quantizations at varying levels and found that $R=40$ produces signals that remain highly recognizable. However, we chose $R=64$ to ensure minimal information loss. 

Once quantized, we processed all the audio in our dataset using a curve length of $L$ samples. We created a dictionary, the key of which was every unique curve sequence that we encountered, and the value of which was the number of times that curve had been seen. At $L=8$, sliding the the $L$-length window across each audio signal with a stride of 1, on ESC-50, this produced a dictionary of $3.87*10^7$ keys. We took the top 50,000 sequences as our vocabulary, which covers 76.49\% of the observed curves. At inference time, we used a stride of $L=8$, which resulted in a overall sequence length decrease of $L$, also. We find that when curve-tokenizing our audio signals in this way, 76.39\% of the curves are found in the vocabulary, with the remaining 23.61\% represented by the equivalent of the \code{<UNK>} token in NLP.

We also created a relative vocabulary by shifting the quantized values, such that the minimum value in any 8-token span was set to zero, and all the other values maintained their relative position to the minimum, according to the Equation \ref{eq:relative-curves}, where $X=\{x_0, x_1, ..., x_n\}$ is a span of audio with individual quantized values $x_i$.

\begin{equation}\label{eq:relative-curves}
X = \sum\limits_{i=0}^{n} x_i - min(X)
\end{equation} 

Using the top 50,000 spans from the relative vocabulary, we find that the it covers 85.44\% of the number of unique spans in the dataset. When using the relative vocabulary to tokenize audio from the dataset, we find that an average 85.43\% of the curves in each audio clip are represented, with 14.57\% represented by the \code{<UNK>} token.

\subsubsection{VQ-VAE}

This method was motivated by Jukebox \cite{dhariwal2020jukebox}, which made use of vector-quantized variational autoencoders to produce compressed ``codes'' to represent audio. The VQ-VAEs that we trained used the code that the authors provided, and details on the specifics of training can be found in their paper. We used their default VQ-VAE hyperparameters, which trained three VQ-VAEs, each with a codebook size of 2048, a total model size of 1 billion parameters, and downsampling rates of 128, 32, and 8. We trained the VQ-VAEs on AudioSet for 500,000 steps. In our experiments, we use the VQ-VAE with a downsampling rate of 32x.

\subsubsection{MFCC}

Mel-frequency cepstral coefficients (MFCCs) have a long history of use in audio classification problems \cite{cowling2003comparison,chachada_environmental_2014,sharma_environment_2020}, and so we tested their usefulness with transformers, as well. Unless otherwise mentioned, we used 128 mels, a hop length of 512, a window length of 1024, and number of FFTs of 1024. 

\subsubsection{MFCC, GFCC, CQT, and Chromagram}

Sharma \et \cite{sharma_environment_2020} reported a new state of the art on ESC-50, using four feature channels at once. They made use of MFCCs, gammatone frequency cepstral coefficients (GFCCs), a constant Q-transform (CQT), and a chromagram. 
Roughly speaking, the usefulness of each feature can be broken down in the following way: MFCCs are responsible for higher-frequency audio, such as speech or laughs; GFCCs are responsible for lower-frequency audio, such as footsteps or drums; CQT is responsible for music; and chromagrams are responsible for differentiating in difficult cases through the use of pitch profiles. A more extended discussion of these features is available in Sharma \et's work \cite{sharma_environment_2020}. We made use of the same features with our transformer models, using the same parameters for feature extraction as Sharma \et. In order to facilitate feeding the features into the transformer model, we concatenate the features, creating a combined feature vector of 512, which became the size of the hidden dimension.

\subsubsection{Mel spectrogram}

Other works obtaining high accuracies on ESC-50, such as the work by Salamon and Bello \cite{salamon_deep_2017}, and, more recently, Kumar \et's work with transfer learning and CNNs \cite{kumar_knowledge_2018}, made use of Mel spectrograms. Therefore, we also chose to include the Mel spectrogram as a feature extraction method.

Motivated by early attempts at downsampling the spectrogram, and seeing little to no decrease in accuracy on ESC-50, we perform downsampling on the spectrogram in order to reduce memory usage, which sped up experiments. The downsampling was performed by taken every $N$th column of the spectrogram matrix, where the column was frequency data at a particular timestep. In our experiments with ESC-50, we used $N=2$ and $N=3$. In our experiments with the Office Sounds dataset, we used $N=1$, or no downsampling. In experiment \#9 on ESC-50 (Table \ref{tab:main-results}), we used 128 Mel bands, 1048 FFTs, hop length of 512, and window length of 1024. In experiment \#10 on ESC-50, we used 256 Mel bands, 2048 FFTs, hop length of 512, and window length of 1024.

\begin{table*}[t]
  \begin{center}
      \caption{Accuracy on ESC-50 dataset, running under various feature extraction and training schemes.}
    \label{tab:main-results}
    
     \begin{tabular*}{0.95\textwidth}{r | l | S | c | c | c | c | c | c | c}
     \toprule
     \# & Input & Accuracy & Samples & Layers & Heads & Sequence Len & Batch & Augment & Type \\
     \midrule
     
     1 & Amplitude reshaping & 48.96 & 44100 & 8 & 8 & 256 & 16 & True & BERT \\ 
     
     2 & Amplitude reshaping (Pretrained) & 52.08 & 44100 & 16 & 16 & 256 & 16 & True & BERT \\ 
     
     3 & VQ-VAE (32x) & 31.77 & 16384 & 8 & 8 & 512 & 32 & False & BERT \\ %
     4 & VQ-VAE (32x) & 34.50 & 65536 & 8 & 8 & 2048 & 4 & False & BERT \\ %
     
     5 & MFCC & 53.13 & 44100 & 8 & 8 & 173 & 32 & False & BERT \\ 
     6 & MFCC & 58.33 & 44100 & 8 & 8 & 173 & 32 & True & BERT \\ 
     
     7 & MFCC, GFCC, CQT, and Chromagram & 59.38 & 88200 & 8 & 8 & 173 & 32 & False & BERT \\ 
     8 & MFCC, GFCC, CQT, and Chromagram & 59.90 & 88200 & 8 & 8 & 173 & 32 & True & BERT \\ 
     
     9 & Mel spectrogram (Downsampled 3x) & 60.45 & 220500 & 8 & 8 & 143 & 64 & False & AlBERT \\ 
     10 & Mel spectrogram (Optimized) & 67.71 & 220500 & 16 & 16 & 215 & 16 & True & AlBERT \\ 
     
     11 & Curve Tokenization (Relative) & 7.81 & 4096 & 8 & 8 & 512 & 16 & False & BERT \\ %
     12 & Curve Tokenization (Relative) & 8.85 & 4096 & 8 & 8 & 512 & 16 & True & BERT \\ %
     
     13 & Curve Tokenization (Absolute) & 19.79 & 4096 & 8 & 8 & 512 & 16 & False & BERT \\
     14 & Curve Tokenization (Absolute) & 13.54 & 4096 & 8 & 8 & 512 & 16 & True & BERT \\ 

     \bottomrule
     \end{tabular*}
  \end{center}
\end{table*}

\subsection{Augmentations}

Inspired by Sharma \et \cite{sharma_environment_2020}, we performed a number of augmentations to the our raw audio. We performed eleven different augmentations:

\begin{itemize}
    \item \textbf{Amplitude clipping}: all samples are clipped at a random amplitude, determined by a percentage range, from 0.75 to 1.0, based on the maximum value in the audio.
    \item \textbf{Volume amplification}: all samples are multiplied by a random value, determined by a percentage range between 0.5 and 1.5.
    \item \textbf{Echo}: a random delay is selected between 2\% and 40\% of one second, and, for each value in the audio, values from the delay value number samples prior to it are added to it. E.g. at index 10,000 in an audio clip, with a random delay number of samples of 4,410, the sample from index 5590 is added to the sample at index 10,000.
    \item \textbf{Lowpass filter}: a fifth-order lowpass filter is passed over the audio, with a cutoff determined by a random value between 0.05 and 0.20.
    \item \textbf{Pitch}: the pitch is shifted by a random value from 0 to 4, using a function provided by librosa, \code{librosa.effects.pitch\_shift}.
    \item \textbf{Partial erase}: a random amount of audio, from 0 to 30\%, is replaced with Gaussian noise.
    \item \textbf{Speed adjust}: The speed of the audio is adjusted randomly between a value of 0.5 and 1.5, where greater than one is faster, and less than one is slower, using a function provided by librosa, \code{librosa.effects.time\_stretch}.
    \item \textbf{Noise}: a random amount of Gaussian noise is added to every sample in the audio.
    \item \textbf{HPSS}: harmonic percussive source separation is performed, with a random choice between returning the harmonic part or the percussive part of the audio.
    \item \textbf{Bitwise downsample}: audio is downsampled by multiplying each sample by a resolution value $R$ between 40 and 100, taking the floor of the value, and then dividing by the resolution. This reduces every sample in the audio to be represented by a maximum of $R$ possible values.
    \item \textbf{Sampling rate downsample}: a value $k$ is selected between 2 and 9, inclusive, and for every audio sample $x_i$, where $i=0,k,2k,...$, the next $k$ positions in the audio are overwritten with $x_i$. The number of samples in the audio stays the same with this method, but the overall information content of the audio is decreased. This method is similar to augmentations that downsample an image, while keeping the size of the image the same.
\end{itemize}

    

\begin{figure*}[ht]
\centering 
    \begin{subfigure}[]{.5\textwidth}
    \centering
    \includegraphics[width=.9\linewidth]{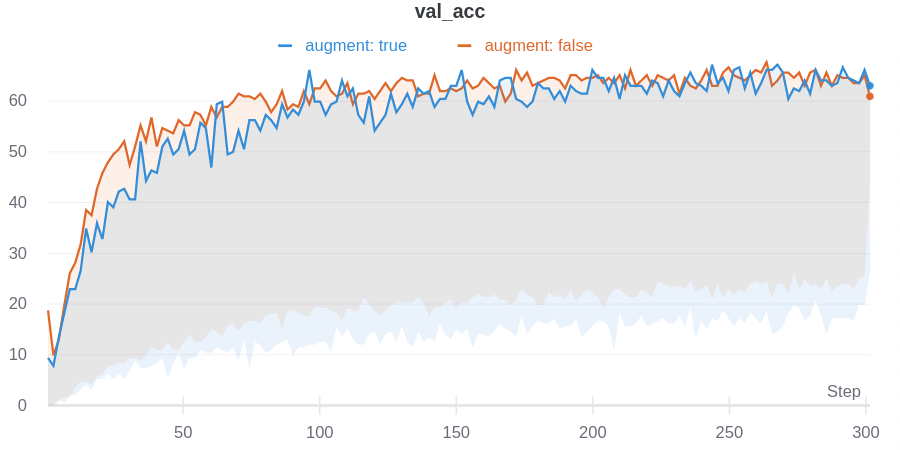}\label{fig:train_acc}
    \captionof{figure}{}
    \end{subfigure}%
    \begin{subfigure}[]{.5\textwidth}
    \centering
    \includegraphics[width=.9\linewidth]{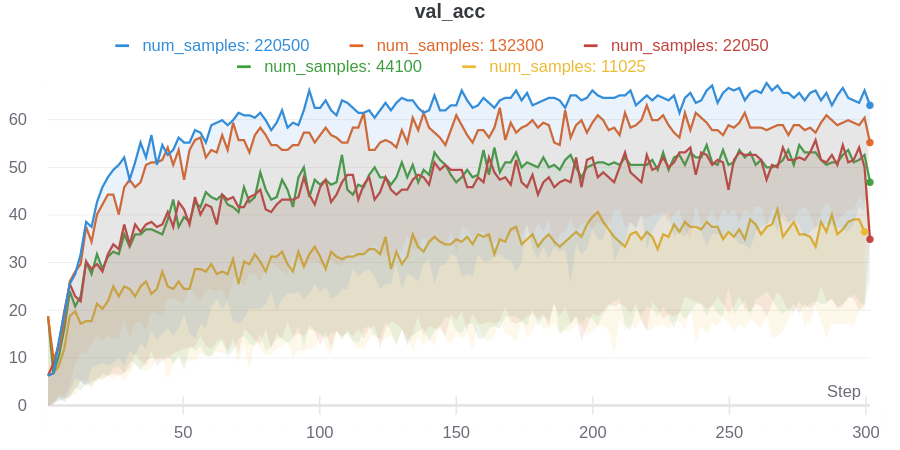}\label{fig:val_acc}
    \captionof{figure}{}
    \end{subfigure}%
    
    \begin{subfigure}[]{.5\textwidth}
    \centering
    \includegraphics[width=.9\linewidth]{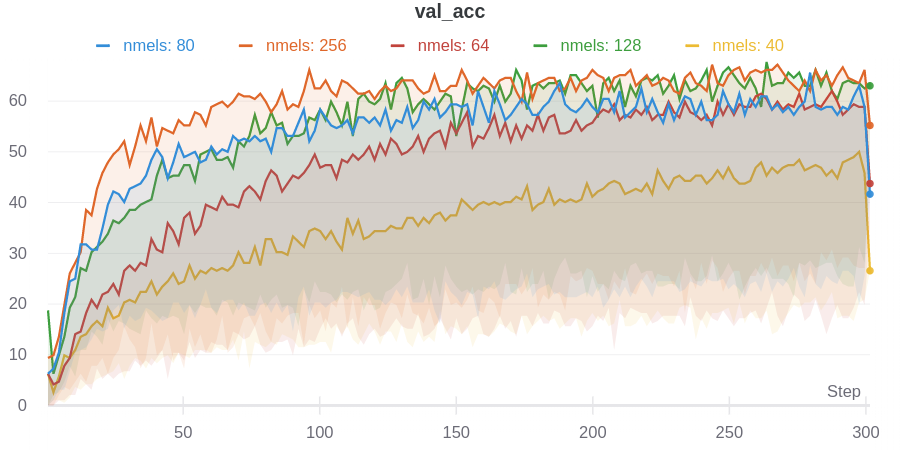}\label{fig:val_acc}
    \captionof{figure}{}
    \end{subfigure}%
    \begin{subfigure}[]{.5\textwidth}
    \centering
    \includegraphics[width=.9\linewidth]{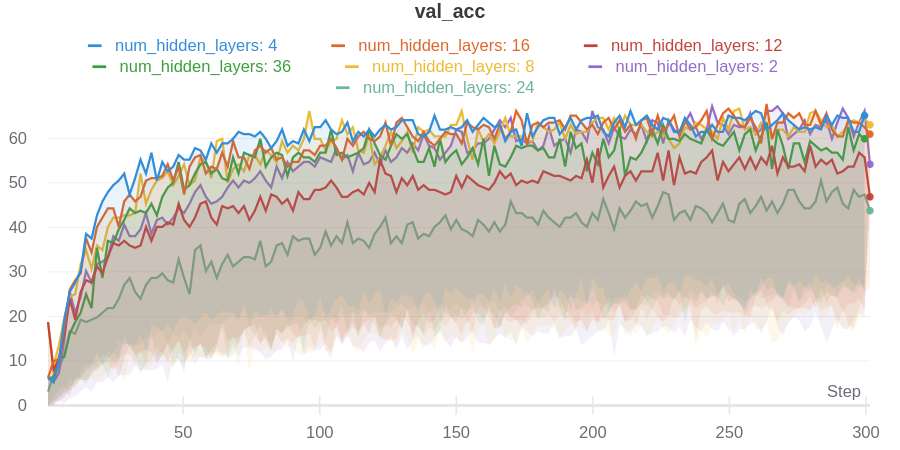}\label{fig:train_acc}
    \captionof{figure}{}
    \end{subfigure}%
    
    \begin{subfigure}[]{.5\textwidth}
    \centering
    \includegraphics[width=.9\linewidth]{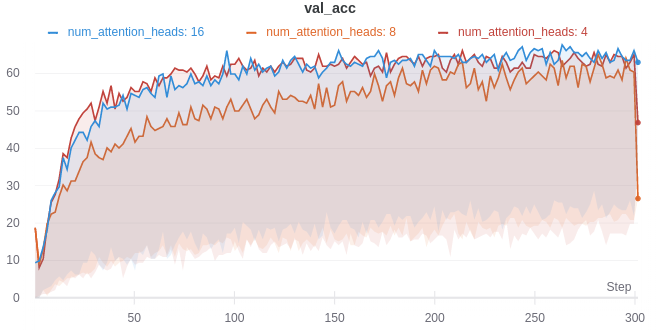}\label{fig:val_acc}
    \captionof{figure}{}
    \end{subfigure}%
    \begin{subfigure}[]{.5\textwidth}
    \centering
    \includegraphics[width=.9\linewidth]{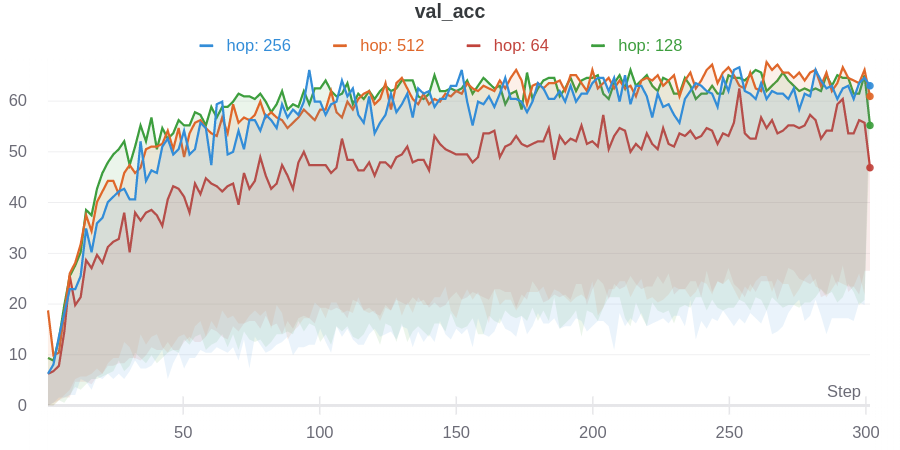}\label{fig:val_acc}
    \captionof{figure}{}
    \end{subfigure}%
  \caption{Validation accuracy on AlBERT, trained using a Mel spectrogram with varying parameters, aggregated over a total of 159 runs. The figures show results as the following parameters are varied: (a) augmentations, (b) number of samples viewed by the model at once, (c) number of Mel bands in the Mel spectrogram, (d) number of hidden layers in the model, or the depth of the model, (e) number of attention heads, and (f) the hop length when calculating the Mel spectrogram.} 
  \label{fig:hyperparameter-search}
\end{figure*}

\subsection{Model Conversion}

To convert the model, we used PyTorch mobile and torchscript \footnote{\url{https://pytorch.org/mobile/home/}}. We also quantized the model using PyTorch's dynamic quantization, which is a part of PyTorch Mobile. 

We did not perform static quantization due to complexity and time constraints. We converted the model into Open Neural Network Exchange (ONNX) format \footnote{\url{https://onnx.ai/}}, in an attempt to convert to TensorFlow, then to TensorFlow Lite. However, we were unsuccessful in this attempt, due to various limitations in the frameworks and conversion process.

Similarly, we attempted to convert the model to a representation that is supported on a Arduino Nano 33 BLE Sense. We attempted to convert the ONNX version of our model to TensorFlow Lite, but encountered multiple issues, one related to missing operators. We also attempted to convert it to deepC \footnote{\url{https://github.com/ai-techsystems/deepC}}, but encountered similar issues, including missing support for quantized PyTorch models. We also did not complete the conversion to a microcontroller-supported representation due to complexity and time constraints.

\section{Experiments}
\label{sec:experiments}

We performed two sets of experiments, one on ESC-50 using the six feature extraction methods described in Section \ref{sec:feature-extraction}, and a second on our Office Sounds dataset, using the best model from the first set of experiments, with some adjustments.

We used HuggingFace's Transformers library \footnote{\url{https://github.com/huggingface/transformers}} for our transformer implementations. Note that HuggingFace's library assumes that a positional embedding is desirable, and has no option to remove it. Therefore, we ran a modified version of their code for our experiments that did not return integer tokens during feature extraction, namely, raw amplitudes, MFCCs, GFCCs, CQTs, Chromagrams, and Mel spectrograms. We did use positional embeddings in our curve-tokenized and VQ-VAE experiments.

We used librosa v0.7.2 \footnote{\url{https://github.com/librosa/librosa}} for Mel spectrogram, MFCC, CQT, and Chromagram feature extraction, and spafe v0.1.2 \footnote{\url{https://github.com/SuperKogito/spafe}} for GFCC feature extraction. We also used Python v3.7.7, PyTorch v1.6.0, and PyTorch Lightning v0.7.6 for machine learning. To make our experiments more accessible, we designed our models to be able to run on consumer hardware. We used two NVIDIA RTX 2080 Ti's to train all of our models, each with 11GB of RAM, with the exception of the VQ-VAE with a sequence length of 2048, experiment \#4, for which we used a NVIDIA Tesla V100 with 16GB of RAM.

We trained using a learning rate of $0.0001$, a learning rate warmup of $10000$ steps, and the Adam optimizer. Our data pipeline is implemented such that, every epoch, a random slice is taken from each audio file, optionally passed through augmentations, and then passed to the model. This has the advantage of vastly simplifying the data processing implementation, and increasing the number of ways in which a model can view a particular sound (assuming that the number of samples viewed by the model is less than the number of samples in the audio file). It does, however, have the disadvantage of reading from every audio file an equal number of times, regardless of the length of the audio. This was not a substantial issue for us, as AudioSet, ESC-50, and Office Sounds all contain roughly the same length audio files within themselves.

\subsection{Experiments on ESC-50}
\label{sec:experiments-esc50}

Table \ref{tab:main-results} describes the results of the trainings that we performed with each of our model types, and we discuss the results below.



\subsubsection{Amplitude Reshaping}

Experiment \#1 with amplitude reshaping tested how well a transformer could learn to predict under a few unusual circumstances: (1) the inputs to the models are not constant with respect to tokens, as is usually the case with learned embeddings, (2) the model is not pretrained, and (3) the dataset is small. The performance of this model was far below comparable CNNs, but better than expected, given that transformers traditionally are pretrained with massive amounts of data, and are known to perform poorly when trained on small datasets alone. We observed that the model began to overfit around 60 epochs.

We also performed a supervised pretraining on reshaped raw amplitudes in experiment \#2. This pretraining comes in the form of training on audio from AudioSet, described in Table \ref{table:data}, which has 527 labels. We trained on AudioSet for 75 epochs, with augmentations, to a maximum top-1 validation accuracy of 6.36\%, after which it began to overfit. We then took that pretrained model, and finetuned it on ESC-50, without freezing any layers, according to standard practice with transformers. It is notable that this pretraining increased accuracy by 3\%, compared to the non-pretrained model. When pretrained on a much larger dataset than AudioSet, it may be the case, as in \cite{anonymous2021an}, that a model like this obtains far higher accuracy when finetuned.

\begin{table*}[t]
  \begin{center}
      \caption{Transformer accuracy on Office Sounds dataset for various models, ordered by number of paramaters. All models were based on BERT, and had the feed-forward layer size set to $4H$.}
    \label{tab:transformers-office-sounds-results}
    
     \begin{tabular*}{\textwidth}{r | l | S | r | r | c | c | c | c | c | c | c}
     \toprule
     \# & Input & Accuracy & Params & Multiply-Adds & Samples & Layers & Hidden & Heads & Sequence Len & Batch & Augment \\
     \midrule
     
     1 & Mel spectrogram & 81.48\% & 5,954 & 5,638 & 44100 & 1 & 16 & 2 & 86 & 64 & True \\
     2 & Mel spectrogram & 93.21\% & 6,642 & 5,982 & 220500 & 1 & 16 & 2 & 430 & 64 & True \\
     3 & Mel spectrogram & 95.31\% & 213,858 & 210,414 & 220500 & 4 & 64 & 4 & 430 & 64 & True \\
     4 & Mel spectrogram & 93.75\% & 25,553,762 & 25,506,734 & 220500 & 8 & 512 & 8 & 430 & 16 & True \\
     5 & Mel spectrogram & 89.38\% & 25,553,762 & 25,506,734 & 220500 & 8 & 512 & 8 & 430 & 16 & False \\

     \bottomrule
     \end{tabular*}
  \end{center}
\end{table*}

\subsubsection{VQ-VAE}

We were surprised by the inneffectiveness of VQ-VAE codes in producing good classifications. Judging by Jukebox \cite{dhariwal_jukebox_2020}, it seemed reasonable to believe that the VQ-VAE would encode a substantial amount of knowledge in the codes, which, if they are enough to produce a good reconstruction of the original audio, might also be enough to produce a classification. We did not find this to be the case, however, as they vastly underperformed compared to MFCCs, raw audio, and others. We can think of several reasons for this: first, the lack of large-scale data reduces the maximum accuracy that can be obtained from any input by a transformer, and this may be particularly true for VQ-VAE codes, since it could have been compounded by the lack of data supplied to both the VQ-VAE in learning codes through AudioSet, and the lack of data in learning classifications in ESC-50. Second, the heterogeneity of sounds in AudioSet may have significantly limited the VQ-VAE's ability to represent sounds in the codes. It has previously been shown that VAEs in general do not perform well on heterogeneous datasets \cite{nazabal2020handling}. As such, our VQ-VAE may not be able to perform as well on environment sound tasks as it did on music tasks, given the large variety of sounds present in ESC versus music. 

Hypothesizing that the short sequence length of our first experiment (512) may have resulted in the transformer not be able to have a sufficient view of the code to perform a classification, we attempted a much longer sequence length, using a V100 GPU with 16GB of RAM. Even with a sequence length of 2048, which translates to an effective number of samples of 65,536, or about 1.5 seconds, we did not observe a substantial increase in accuracy, still falling far below other feature extraction methods.

As with the rest of the methods in this work, the first step to increasing accuracy on ESC using VQ-VAE codes is to obtain more data. Training on a much larger corpus of unlabeled audio is entirely possible in the first step to creating VQ-VAE codes, and may improve the quality of the codes created. Additionally, using techniques such as the ones presented by Nazabal \et \cite{nazabal2020handling}, to alter the VQ-VAE to enable to it better handle heterogeneous data, may help as well. It may also be of value to perform a pretraining step, either supervised or unsupervised, and finetune on more ESC data. However, even with all such adjustments, it seems unlikely that VQ-VAE codes will exceed MFCCs, Mel spectrograms or raw audio in predictive capability.

\subsubsection{MFCC, GFCC, CQT, and Chromagram}
In experiments \#5 and \#6, we observe that augmentations make a substantial (5.2\%) impact on accuracy. We also see that MFCC's perform better, though only slightly so, than raw amplitudes. These experiments were performed with 128 Mel bands, which resulted in the hidden size $H$ of the model to be 128 as well. These models began to overfit around 50 epochs. 

Experiments \#7 and \#8 showed that adding additional feature extraction methods improved the accuracy of the model beyond only using MFCCs, especially in the non-augmented case. However, when augmented, the model did not show any major improvements, as had occurred with MFCCs. This is different than the results by Sharma \et \cite{sharma_environment_2020}, which had used augmentations to improve accuracy by more than 3\%. However, we note that for our purposes -- inferring at the edge -- the cost of computing features using all four extraction methods becomes prohibitive, and the model would have been unlikely to be of use at the edge, even it it had obtained high accuracy. We also found that extracting these features at training time resulted in extremely slow training, which hindered additional experimentation with these features.

\begin{table*}[h]
  \begin{center}
      \caption{CNN accuracy on Office Sounds dataset, ordered by accuracy.}
    \label{tab:mfcc-office-sounds}
    
     \begin{tabular*}{0.63\textwidth}{r | l | S | r | r | c | c | c}
     \toprule
     \# & Input & Accuracy & Params & Multiply-Adds & Samples & Batch & Augment \\
     
     \midrule
     
     1 & MFCC & 92.97\% & 4,468,022 & 478,869,984 & 110250 & 64 & True \\
     2 & MFCC & 92.19\% & 4,468,022 & 478,869,984 & 110250 & 64 & False \\
     3 & MFCC & 91.41\% & 4,468,022 & 956,052,832 & 220500 & 16 & False \\
     
     \bottomrule
     \end{tabular*}
  \end{center}
\end{table*}

\subsubsection{Mel Spectrogram and Hyperparameter Search}

We trained using a Mel spectrogram in experiments \#9 and \#10, and obtained accuracy that outperformed any other feature extraction methods. This is particularly advantageous at the edge, since computing a Mel spectrogram is a reasonably inexpensive operation. Of note, as well, is the fact that this was obtained with a smaller sequence length than experiments \#7 and \#8, due to the 3x downsampling that we performed. We also used AlBERT, and a longer sequence length that other models, which may have contributed to the improved performance. Judging by the performance of BERT-based transformers trained on Office Sounds, however, it seems unlikely that AlBERT would result in a significant performance improvement alone. The impact of number of samples is discussed below.

Since this was our best-performing model, we performed a hyperparameter search to determine the optimal parameters. All training was performed using AlBERT as the base model, with a downsampling rate of 2x. We performed 159 training runs, which are aggregated into the graphs in Figure \ref{fig:hyperparameter-search}.

Some clear improvements result by changing certain parameters. The most obvious is the number of samples that are passed into the Mel spectrogram, which, as it increases, also increases the maximum possible validation accuracy. We chose a peak of 220,500 samples, or 5 seconds of audio, because files in ESC-50 audio have a maximum length of 5 seconds. As can be seen, a model's access to the full file's worth of data improves its ability to classify well.

Another clear result is the importance of using more than 80 Mel bands when creating the Mel spectrogram. This result is particularly important, as many research works make use of 80 Mels or less \cite{piczak_environmental_2015,li_comparison_2017,zhao2020musicoder,zhang_transformer_2020,miyazaki_weakly-supervised_2020,jiao2020translate}, which likely reduced accuracy in those works.

\subsubsection{Curve Tokenization}

As a first attempt in literature at tokenizing audio based on curves, for the purpose of training a transformer, we find that they provide very little predictive power. There may be several reasons for this, the first of which is the small number of samples over which the model can view a sound. Since every 8 samples is quantized and converted into a token, using a sequence length of 512, the number of effective samples is 4096, which is only 93 milliseconds of audio. This is likely a limiting factor on the predictive ability of the model, and a model able to handle a much longer sequence length would likely perform better. It is also likely that quantizing it reduced the information content of the audio, and further reduced the predictive power.

In the case of absolute curves, we find that augmentations substantially reduce accuracy. This is likely due to the fact that our vocabulary was created on ESC-50 without augmentations, so the curves that appear with augmentations result in many more \code{<UNK>} tokens. We see a slight increase in relative curve tokenization with augmentations, but, given the incredibly low accuracy of the model, find it to be of little interest.

Overall, we consider it unlikely that curve tokenization would ever beat out more well-known feature extraction techniques. It removes too much information, such as vital frequency and phase information, which other feature extraction methods allow the transformer to make use of. Nonetheless, we consider it an interesting experiment in possible tokenization techniques for transformers on audio.

\subsection{Experiments on Office Sounds}
\label{sec:experiments-officesounds}

After completing our experiments on ESC-50, we trained on the Office Sounds dataset \cite{elliott2020cyber}. We used BERT-based models only, with an emphasis on model size, specifically on reducing the model size while maintaining accuracy in order to perform more efficient processing at the edge. We were particularly interested in models which were capable of being run on microcontrollers; in our case, we chose a target model size of 256KB or less -- the available SRAM on the Arduino Nano 33 BLE Sense -- which meant that the model must be less than 250,000 parameters when quantized. There are, of course, methods to run larger models with less SRAM, such as MCUNet \cite{lin2020mcunet}, but we left such optimizations to a future work, focusing on the generic case of running a transformer on a microcontroller without any special optimizations.

We made several adjustments between our ESC-50 experiments and our Office Sounds experiments, described in Section \ref{sec:models}, which enabled us to experiment with vastly different model sizes. We began by choosing a model with parameters similar to our best-performing models from the hyperparameter search. We chose the model seen in experiments \#4 and \#5 of Table \ref{tab:transformers-office-sounds-results}, based on Mel spectrogram input with a hop length of 512, window size of 1024, number of FFTs of 1024, and Mel bands of 128. It had 8 layers, and a hidden size $H$ larger than we had been able to use in the ESC-50 experiments, of 512, and 8 heads. We also removed downsampling, making the sequence length much longer than before, but still able to fit within the constraints of consumer-grade GPUs while maintaining a reasonable batch size. These models obtained a maximum validation accuracy of 93.75\%, with augmentations, and began to overfit after 200-300 epochs.

In order to facilitate accurate comparisons to our previous work \cite{elliott2020cyber}, we reimplemented and performed training on Office Sounds using MFCCs as input to a CNN, shown in Table \ref{tab:mfcc-office-sounds}. Following that work, the CNN was an exact reimplementation of Kumar \et's model in \cite{kumar_knowledge_2018}. We trained the model, non-augmented, on ESC-50 to confirm accurate implementation, and obtained 81.25\%, which is very close to the 83.5\% model accuracy that Kumar \et reported for their model that had been pretrained on AudioSet. We performed training of this MFCC-based CNN against Office Sounds, using a random slice of the each audio file in each epoch, and obtained a maximum of 92.97\% accuracy, using augmentations on 2.5 seconds of audio. This corresponds to the results obtained in \cite{elliott2020cyber}, even though the training scheme is slightly different. The model contained 4.5 million parameters, and nearly 500 million multiply-adds. The inference time of this model on a Samsung Galaxy S9 was an average of 57 milliseconds \cite{elliott2020cyber}, non-augmented and non-quantized, using TensorFlow Lite, shown in Table \ref{tab:mobile-inference}. We note that augmentations had a small positive effect on validation accuracy, and that increasing the visible audio window size from 2.5 to 5 seconds had a slightly negative effect.

In comparing the Office Sounds transformers to the Office Sounds CNNs, we find that the transformers outperform the CNN, while being much smaller. A model with 95.2\% less parameters, transformer experiment \#3, outperformed the CNN by more than 2\%. The smallest model that we trained on 5 seconds of audio, experiment \#2, 99.85\% smaller than the CNN, also outperformed the CNNs. This is unexpected, since CNNs far outperformed transformers on the ESC-50 experiments. Our hypothesis is that the increased number of example data for each class in Office Sounds (200 or more per class, compared to 40 per class in ESC-50), assisted in preventing as rapid overfitting as was observed in ESC-50. This can be tested by reducing the number of samples in Office Sounds and running these experiments again; we leave this for a future work.

We also trained our smallest model on one second of data (experiment \#1), and found that it substantially reduced the accuracy of the model. Interestingly, for some applications, it may be worthwhile to process only one second of audio with reduced accuracy, as it reduces the cost of feature extraction and allows the model to be run more frequently. Also, predictions over each second can be aggregated across a longer time span via majority voting, or something similar, in order to potentially produce more accurate predictions.

\subsection{Inference at the Edge}

Table \ref{tab:mobile-inference} shows feature extraction and inference times on a Samsung Galaxy S9. This model uses a transformer based on a Mel spectrogram, processing 5 seconds of audio data and producing a classification using ESC-50 labels. We observe that, even on a device more than two years old, inference is still fast enough to be performed many times a second. We also find that quantization results in lowered latency (about 21\% with dynamic quantization), which further increases the potential model size. Static quantization is likely to reduce latency further, as dynamic quantization does not quantize model activations.

\begin{table}[h!]
  \begin{center}
    \caption{Inference times of selected models on an edge device. Models are run on a Samsung Galaxy S9 using PyTorch Mobile, except for the first, which was run with TensorFlow Lite. Results are averaged over 10 runs. Quantization is PyTorch dynamic quantization.}
    \label{tab:mobile-inference}
    
    \begin{tabular}{l|c|c|l}
      \toprule 
      \textbf{Experiment} & \textbf{Params} & \textbf{Mult-Adds} & \textbf{Latency (ms)} \\
      \midrule 
      MFCC CNN from \cite{elliott2020cyber} & 4,468,022 & 478,869,984 & 57 \\ 
      ESC-50 \#10 & 958,146 & 948,888 & 111 \\ 
      ESC-50 \#10, Quant. & 958,146 & 948,888 & 88 \\ 
      Office Sounds, Trans. \#2 & 6,642 & 5,982 & 7 \\
      \bottomrule 
    \end{tabular}
  \end{center}
\end{table}

We also observed a substantially decreased inference time from our smallest model, as expected, inferring 93\% faster than the 1-million parameter transformer. It was surprising to find that the much larger CNN had faster inference times than the 1-million parameter transformer, however, this may be due to optimizations in TensorFlow Lite that are not present in PyTorch Mobile, or simply that CNN operations are optimized further than transformer operations on edge devices.


\section{Conclusion and Future Work}

Efficient edge processing is a challenging, but critical task which will become increasingly important in the future. To aide in this task, we trained a 6,000-parameter transformer on the Office Sounds dataset that outperforms a CNN more than 700x larger than it. This enables accurate and efficient environmental sound classification of office sounds on edge devices, even on inexpensive microcontrollers, resulting in inference times on a Samsung Galaxy S9 that are 88\% faster than a CNN with comparable accuracy. We find that models trained in traditional frameworks (like PyTorch) have relatively little support for conversion to models that can be run at the edge (like on a microcontroller), even with the development of ONNX.

Our ESC-50 transformer models did not outperform CNNs, as they did on Office Sounds. Understanding this, and finding solutions to the problem of training transformers on small audio datasets, is a crucial future work. Solutions may come through large amounts of unsupervised pretraining, through an architectural change, or though improved supervised ESC datasets. In any case, our work provides groundwork upon which these questions can be answered.

The small size and efficiency of the transformer we trained raises questions about the cost of retraining. It may be that, because there are so few operations ($<$6,000) required in a forward pass, that \textit{on-device} retraining becomes possible, similar to what is done on Coral Edge TPUs through the imprinting engine \cite{coral}. This would have vast implications on the future of intelligent edge analytics, and even a variety of user applications.





\bibliographystyle{IEEEtran}
\bibliography{main}

\begin{thebibliography}{10}
\providecommand{\url}[1]{#1}
\csname url@samestyle\endcsname
\providecommand{\newblock}{\relax}
\providecommand{\bibinfo}[2]{#2}
\providecommand{\BIBentrySTDinterwordspacing}{\spaceskip=0pt\relax}
\providecommand{\BIBentryALTinterwordstretchfactor}{4}
\providecommand{\BIBentryALTinterwordspacing}{\spaceskip=\fontdimen2\font plus
\BIBentryALTinterwordstretchfactor\fontdimen3\font minus
  \fontdimen4\font\relax}
\providecommand{\BIBforeignlanguage}[2]{{%
\expandafter\ifx\csname l@#1\endcsname\relax
\typeout{** WARNING: IEEEtran.bst: No hyphenation pattern has been}%
\typeout{** loaded for the language `#1'. Using the pattern for}%
\typeout{** the default language instead.}%
\else
\language=\csname l@#1\endcsname
\fi
#2}}
\providecommand{\BIBdecl}{\relax}
\BIBdecl

\bibitem{cowling2003comparison}
M.~Cowling and R.~Sitte, ``Comparison of techniques for environmental sound
  recognition,'' \emph{Pattern recognition letters}, vol.~24, no.~15, pp.
  2895--2907, 2003.

\bibitem{vaswani2017attention}
A.~Vaswani, N.~Shazeer, N.~Parmar, J.~Uszkoreit, L.~Jones, A.~N. Gomez,
  {\L}.~Kaiser, and I.~Polosukhin, ``Attention is all you need,'' in
  \emph{Advances in neural information processing systems}, 2017, pp.
  5998--6008.

\bibitem{devlin2018bert}
J.~Devlin, M.-W. Chang, K.~Lee, and K.~Toutanova, ``Bert: Pre-training of deep
  bidirectional transformers for language understanding,'' \emph{arXiv preprint
  arXiv:1810.04805}, 2018.

\bibitem{yang2019xlnet}
Z.~Yang, Z.~Dai, Y.~Yang, J.~Carbonell, R.~R. Salakhutdinov, and Q.~V. Le,
  ``Xlnet: Generalized autoregressive pretraining for language understanding,''
  in \emph{Advances in neural information processing systems}, 2019, pp.
  5753--5763.

\bibitem{raffel2019exploring}
C.~Raffel, N.~Shazeer, A.~Roberts, K.~Lee, S.~Narang, M.~Matena, Y.~Zhou,
  W.~Li, and P.~J. Liu, ``Exploring the limits of transfer learning with a
  unified text-to-text transformer,'' \emph{arXiv preprint arXiv:1910.10683},
  2019.

\bibitem{radford2018improving}
A.~Radford, K.~Narasimhan, T.~Salimans, and I.~Sutskever, ``Improving language
  understanding by generative pre-training,'' 2018.

\bibitem{radford2019language}
A.~Radford, J.~Wu, R.~Child, D.~Luan, D.~Amodei, and I.~Sutskever, ``Language
  models are unsupervised multitask learners,'' \emph{OpenAI Blog}, vol.~1,
  no.~8, p.~9, 2019.

\bibitem{brown2020language}
T.~B. Brown, B.~Mann, N.~Ryder, M.~Subbiah, J.~Kaplan, P.~Dhariwal,
  A.~Neelakantan, P.~Shyam, G.~Sastry, A.~Askell \emph{et~al.}, ``Language
  models are few-shot learners,'' \emph{arXiv preprint arXiv:2005.14165}, 2020.

\bibitem{zaheer2020big}
M.~Zaheer, G.~Guruganesh, A.~Dubey, J.~Ainslie, C.~Alberti, S.~Ontanon,
  P.~Pham, A.~Ravula, Q.~Wang, L.~Yang \emph{et~al.}, ``Big bird: Transformers
  for longer sequences,'' \emph{arXiv preprint arXiv:2007.14062}, 2020.

\bibitem{chen2020generative}
M.~Chen, A.~Radford, R.~Child, J.~Wu, H.~Jun, P.~Dhariwal, D.~Luan, and
  I.~Sutskever, ``Generative pretraining from pixels,'' in \emph{Proceedings of
  the 37th International Conference on Machine Learning}, 2020.

\bibitem{dhariwal2020jukebox}
P.~Dhariwal, H.~Jun, C.~Payne, J.~W. Kim, A.~Radford, and I.~Sutskever,
  ``Jukebox: A generative model for music,'' \emph{arXiv preprint
  arXiv:2005.00341}, 2020.

\bibitem{van2017neural}
A.~Van Den~Oord, O.~Vinyals \emph{et~al.}, ``Neural discrete representation
  learning,'' in \emph{Advances in Neural Information Processing Systems},
  2017, pp. 6306--6315.

\bibitem{anonymous2021an}
\BIBentryALTinterwordspacing
Anonymous, ``An image is worth 16x16 words: Transformers for image recognition
  at scale,'' in \emph{Submitted to International Conference on Learning
  Representations}, 2021, under review. [Online]. Available:
  \url{https://openreview.net/forum?id=YicbFdNTTy}
\BIBentrySTDinterwordspacing

\bibitem{pham_very_2019}
\BIBentryALTinterwordspacing
N.-Q. Pham, T.-S. Nguyen, J.~Niehues, M.~Müller, S.~Stüker, and A.~Waibel,
  ``Very deep self-attention networks for end-to-end speech recognition.''
  [Online]. Available: \url{http://arxiv.org/abs/1904.13377}
\BIBentrySTDinterwordspacing

\bibitem{zhang_transformer_2020}
\BIBentryALTinterwordspacing
R.~Zhang, H.~Wu, W.~Li, D.~Jiang, W.~Zou, and X.~Li, ``Transformer based
  unsupervised pre-training for acoustic representation learning,''
  \emph{arXiv:2007.14602 [cs, eess]}, Jul. 2020, arXiv: 2007.14602. [Online].
  Available: \url{http://arxiv.org/abs/2007.14602}
\BIBentrySTDinterwordspacing

\bibitem{shi_weak-attention_2020}
\BIBentryALTinterwordspacing
Y.~Shi, Y.~Wang, C.~Wu, C.~Fuegen, F.~Zhang, D.~Le, C.-F. Yeh, and M.~L.
  Seltzer, ``\BIBforeignlanguage{en}{Weak-{Attention} {Suppression} {For}
  {Transformer} {Based} {Speech} {Recognition}},''
  \emph{\BIBforeignlanguage{en}{arXiv:2005.09137 [cs, eess]}}, May 2020, arXiv:
  2005.09137. [Online]. Available: \url{http://arxiv.org/abs/2005.09137}
\BIBentrySTDinterwordspacing

\bibitem{stowell_dcase_2015}
\BIBentryALTinterwordspacing
D.~Stowell, D.~Giannoulis, E.~Benetos, M.~Lagrange, and M.~D. Plumbley,
  ``\BIBforeignlanguage{en}{{DCASE} 2016 {Acoustic} {Scene} {Classification}
  {Using} {Convolutional} {Neural} {Networks}},''
  \emph{\BIBforeignlanguage{en}{IEEE Transactions on Multimedia}}, vol.~17,
  no.~10, pp. 1733--1746, Oct. 2015. [Online]. Available:
  \url{http://ieeexplore.ieee.org/document/7100934/}
\BIBentrySTDinterwordspacing

\bibitem{gemmeke2017audio}
J.~F. Gemmeke, D.~P. Ellis, D.~Freedman, A.~Jansen, W.~Lawrence, R.~C. Moore,
  M.~Plakal, and M.~Ritter, ``Audio set: An ontology and human-labeled dataset
  for audio events,'' in \emph{2017 IEEE International Conference on Acoustics,
  Speech and Signal Processing (ICASSP)}.\hskip 1em plus 0.5em minus
  0.4em\relax IEEE, 2017, pp. 776--780.

\bibitem{dai_very_2017}
W.~Dai, C.~Dai, S.~Qu, J.~Li, and S.~Das, ``Very deep convolutional neural
  networks for raw waveforms,'' in \emph{2017 {IEEE} International Conference
  on Acoustics, Speech and Signal Processing ({ICASSP})}, pp. 421--425, {ISSN}:
  2379-190X.

\bibitem{salamon_deep_2017}
\BIBentryALTinterwordspacing
J.~Salamon and J.~P. Bello, ``\BIBforeignlanguage{en}{Deep {Convolutional}
  {Neural} {Networks} and {Data} {Augmentation} for {Environmental} {Sound}
  {Classification}},'' \emph{\BIBforeignlanguage{en}{IEEE Signal Processing
  Letters}}, vol.~24, no.~3, pp. 279--283, Mar. 2017. [Online]. Available:
  \url{http://ieeexplore.ieee.org/document/7829341/}
\BIBentrySTDinterwordspacing

\bibitem{tokozume_learning_2017}
Y.~Tokozume and T.~Harada, ``Learning environmental sounds with end-to-end
  convolutional neural network,'' in \emph{2017 {IEEE} International Conference
  on Acoustics, Speech and Signal Processing ({ICASSP})}, pp. 2721--2725,
  {ISSN}: 2379-190X.

\bibitem{zhang_dilated_2017}
X.~Zhang, Y.~Zou, and W.~Shi, ``Dilated convolution neural network with
  {LeakyReLU} for environmental sound classification,'' in \emph{2017 22nd
  International Conference on Digital Signal Processing ({DSP})}, pp. 1--5,
  {ISSN}: 2165-3577.

\bibitem{abdoli_end--end_2019}
\BIBentryALTinterwordspacing
S.~Abdoli, P.~Cardinal, and A.~Lameiras~Koerich,
  ``\BIBforeignlanguage{en}{End-to-end environmental sound classification using
  a {1D} convolutional neural network},'' \emph{\BIBforeignlanguage{en}{Expert
  Systems with Applications}}, vol. 136, pp. 252--263, Dec. 2019. [Online].
  Available:
  \url{https://linkinghub.elsevier.com/retrieve/pii/S0957417419304403}
\BIBentrySTDinterwordspacing

\bibitem{khamparia_sound_2019}
A.~Khamparia, D.~Gupta, N.~G. Nguyen, A.~Khanna, B.~Pandey, and P.~Tiwari,
  ``Sound classification using convolutional neural network and tensor deep
  stacking network,'' vol.~7, pp. 7717--7727, conference Name: {IEEE} Access.

\bibitem{sharma_environment_2020}
\BIBentryALTinterwordspacing
J.~Sharma, O.-C. Granmo, and M.~Goodwin, ``\BIBforeignlanguage{en}{Environment
  {Sound} {Classification} using {Multiple} {Feature} {Channels} and
  {Attention} based {Deep} {Convolutional} {Neural} {Network}},''
  \emph{\BIBforeignlanguage{en}{arXiv:1908.11219 [cs, eess, stat]}}, Apr. 2020,
  arXiv: 1908.11219. [Online]. Available: \url{http://arxiv.org/abs/1908.11219}
\BIBentrySTDinterwordspacing

\bibitem{kumar_knowledge_2018}
\BIBentryALTinterwordspacing
A.~Kumar, M.~Khadkevich, and C.~Fugen, ``\BIBforeignlanguage{en}{Knowledge
  {Transfer} from {Weakly} {Labeled} {Audio} {Using} {Convolutional} {Neural}
  {Network} for {Sound} {Events} and {Scenes}},'' in
  \emph{\BIBforeignlanguage{en}{2018 {IEEE} {International} {Conference} on
  {Acoustics}, {Speech} and {Signal} {Processing} ({ICASSP})}}.\hskip 1em plus
  0.5em minus 0.4em\relax Calgary, AB: IEEE, Apr. 2018, pp. 326--330. [Online].
  Available: \url{https://ieeexplore.ieee.org/document/8462200/}
\BIBentrySTDinterwordspacing

\bibitem{zhang_learning_2019}
\BIBentryALTinterwordspacing
Z.~Zhang, S.~Xu, S.~Zhang, T.~Qiao, and S.~Cao,
  ``\BIBforeignlanguage{en}{Learning {Attentive} {Representations} for
  {Environmental} {Sound} {Classification}},''
  \emph{\BIBforeignlanguage{en}{IEEE Access}}, vol.~7, pp. 130\,327--130\,339,
  2019. [Online]. Available:
  \url{https://ieeexplore.ieee.org/document/8823934/}
\BIBentrySTDinterwordspacing

\bibitem{piczak2015esc}
K.~J. Piczak, ``Esc: Dataset for environmental sound classification,'' in
  \emph{Proceedings of the 23rd ACM international conference on
  Multimedia}.\hskip 1em plus 0.5em minus 0.4em\relax ACM, 2015, pp.
  1015--1018.

\bibitem{salamon2014dataset}
J.~Salamon, C.~Jacoby, and J.~P. Bello, ``A dataset and taxonomy for urban
  sound research,'' in \emph{Proceedings of the 22nd ACM international
  conference on Multimedia}.\hskip 1em plus 0.5em minus 0.4em\relax ACM, 2014,
  pp. 1041--1044.

\bibitem{giannoulis_2013}
D.~{Giannoulis}, E.~{Benetos}, D.~{Stowell}, M.~{Rossignol}, M.~{Lagrange}, and
  M.~D. {Plumbley}, ``Detection and classification of acoustic scenes and
  events: An ieee aasp challenge,'' in \emph{2013 IEEE Workshop on Applications
  of Signal Processing to Audio and Acoustics}, Oct 2013, pp. 1--4.

\bibitem{brown_language_2020}
\BIBentryALTinterwordspacing
T.~B. Brown, B.~Mann, N.~Ryder, M.~Subbiah, J.~Kaplan, P.~Dhariwal,
  A.~Neelakantan, P.~Shyam, G.~Sastry, A.~Askell, S.~Agarwal, A.~Herbert-Voss,
  G.~Krueger, T.~Henighan, R.~Child, A.~Ramesh, D.~M. Ziegler, J.~Wu,
  C.~Winter, C.~Hesse, M.~Chen, E.~Sigler, M.~Litwin, S.~Gray, B.~Chess,
  J.~Clark, C.~Berner, S.~McCandlish, A.~Radford, I.~Sutskever, and D.~Amodei,
  ``\BIBforeignlanguage{en}{Language {Models} are {Few}-{Shot} {Learners}},''
  \emph{\BIBforeignlanguage{en}{arXiv:2005.14165 [cs]}}, May 2020, arXiv:
  2005.14165. [Online]. Available: \url{http://arxiv.org/abs/2005.14165}
\BIBentrySTDinterwordspacing

\bibitem{sanh_distilbert_2020}
\BIBentryALTinterwordspacing
V.~Sanh, L.~Debut, J.~Chaumond, and T.~Wolf,
  ``\BIBforeignlanguage{en}{{DistilBERT}, a distilled version of {BERT}:
  smaller, faster, cheaper and lighter},''
  \emph{\BIBforeignlanguage{en}{arXiv:1910.01108 [cs]}}, Feb. 2020, arXiv:
  1910.01108. [Online]. Available: \url{http://arxiv.org/abs/1910.01108}
\BIBentrySTDinterwordspacing

\bibitem{sun_patient_2019}
\BIBentryALTinterwordspacing
S.~Sun, Y.~Cheng, Z.~Gan, and J.~Liu, ``Patient knowledge distillation for
  {BERT} model compression.'' [Online]. Available:
  \url{http://arxiv.org/abs/1908.09355}
\BIBentrySTDinterwordspacing

\bibitem{turc2019well}
I.~Turc, M.-W. Chang, K.~Lee, and K.~Toutanova, ``Well-read students learn
  better: On the importance of pre-training compact models,'' \emph{arXiv
  preprint arXiv:1908.08962}, 2019.

\bibitem{couvreur_automatic_nodate}
C.~Couvreur, V.~Fontaine, P.~Gaunard, and C.~G. Mubikangiey, ``Automatic
  classification of environmental noise events by hidden markov models,''
  p.~20.

\bibitem{mushtaq172spectral}
Z.~Mushtaq, S.-F. Su, and Q.-V. Tran, ``Spectral images based environmental
  sound classification using cnn with meaningful data augmentation,''
  \emph{Applied Acoustics}, vol. 172, p. 107581.

\bibitem{dhariwal_jukebox_2020}
\BIBentryALTinterwordspacing
P.~Dhariwal, H.~Jun, C.~Payne, J.~W. Kim, A.~Radford, and I.~Sutskever,
  ``Jukebox: A generative model for music.'' [Online]. Available:
  \url{http://arxiv.org/abs/2005.00341}
\BIBentrySTDinterwordspacing

\bibitem{miyazaki_weakly-supervised_2020}
K.~Miyazaki, T.~Komatsu, T.~Hayashi, S.~Watanabe, T.~Toda, and K.~Takeda,
  ``Weakly-supervised sound event detection with self-attention,'' in
  \emph{{ICASSP} 2020 - 2020 {IEEE} International Conference on Acoustics,
  Speech and Signal Processing ({ICASSP})}, pp. 66--70, {ISSN}: 2379-190X.

\bibitem{mitrovic2010features}
D.~Mitrovi{\'c}, M.~Zeppelzauer, and C.~Breiteneder, ``Features for
  content-based audio retrieval,'' in \emph{Advances in computers}.\hskip 1em
  plus 0.5em minus 0.4em\relax Elsevier, 2010, vol.~78, pp. 71--150.

\bibitem{boddapati_classifying_2017}
\BIBentryALTinterwordspacing
V.~Boddapati, A.~Petef, J.~Rasmusson, and L.~Lundberg,
  ``\BIBforeignlanguage{en}{Classifying environmental sounds using image
  recognition networks},'' \emph{\BIBforeignlanguage{en}{Procedia Computer
  Science}}, vol. 112, pp. 2048--2056, 2017. [Online]. Available:
  \url{https://linkinghub.elsevier.com/retrieve/pii/S1877050917316599}
\BIBentrySTDinterwordspacing

\bibitem{piczak_environmental_2015}
K.~J. Piczak, ``Environmental sound classification with convolutional neural
  networks,'' in \emph{2015 {IEEE} 25th International Workshop on Machine
  Learning for Signal Processing ({MLSP})}, pp. 1--6, {ISSN}: 2378-928X.

\bibitem{li_comparison_2017}
\BIBentryALTinterwordspacing
J.~Li, W.~Dai, F.~Metze, S.~Qu, and S.~Das, ``\BIBforeignlanguage{en}{A
  comparison of {Deep} {Learning} methods for environmental sound detection},''
  in \emph{\BIBforeignlanguage{en}{2017 {IEEE} {International} {Conference} on
  {Acoustics}, {Speech} and {Signal} {Processing} ({ICASSP})}}.\hskip 1em plus
  0.5em minus 0.4em\relax New Orleans, LA: IEEE, Mar. 2017, pp. 126--130.
  [Online]. Available: \url{http://ieeexplore.ieee.org/document/7952131/}
\BIBentrySTDinterwordspacing

\bibitem{shankar_novel_2017}
\BIBentryALTinterwordspacing
R.~N. Tak, D.~M. Agrawal, and H.~A. Patil, ``\BIBforeignlanguage{en}{Novel
  {Phase} {Encoded} {Mel} {Filterbank} {Energies} for {Environmental} {Sound}
  {Classification}},'' in \emph{\BIBforeignlanguage{en}{Pattern {Recognition}
  and {Machine} {Intelligence}}}, B.~U. Shankar, K.~Ghosh, D.~P. Mandal, S.~S.
  Ray, D.~Zhang, and S.~K. Pal, Eds.\hskip 1em plus 0.5em minus 0.4em\relax
  Cham: Springer International Publishing, 2017, vol. 10597, pp. 317--325.
  [Online]. Available:
  \url{http://link.springer.com/10.1007/978-3-319-69900-4_40}
\BIBentrySTDinterwordspacing

\bibitem{agrawal_novel_2017}
\BIBentryALTinterwordspacing
D.~M. Agrawal, H.~B. Sailor, M.~H. Soni, and H.~A. Patil,
  ``\BIBforeignlanguage{en}{Novel {TEO}-based {Gammatone} features for
  environmental sound classification},'' in \emph{\BIBforeignlanguage{en}{2017
  25th {European} {Signal} {Processing} {Conference} ({EUSIPCO})}}.\hskip 1em
  plus 0.5em minus 0.4em\relax Kos, Greece: IEEE, Aug. 2017, pp. 1809--1813.
  [Online]. Available: \url{http://ieeexplore.ieee.org/document/8081521/}
\BIBentrySTDinterwordspacing

\bibitem{mogi_noise-robust_2012}
\BIBentryALTinterwordspacing
R.~Mogi and H.~Kasai, ``\BIBforeignlanguage{en}{Noise-{Robust} environmental
  sound classification method based on combination of {ICA} and {MP}
  features},'' \emph{\BIBforeignlanguage{en}{Artificial Intelligence
  Research}}, vol.~2, no.~1, p. p107, Nov. 2012. [Online]. Available:
  \url{http://www.sciedu.ca/journal/index.php/air/article/view/1399}
\BIBentrySTDinterwordspacing

\bibitem{chachada_environmental_2014}
\BIBentryALTinterwordspacing
S.~Chachada and C.-C.~J. Kuo, ``\BIBforeignlanguage{en}{Environmental sound
  recognition: a survey},'' \emph{\BIBforeignlanguage{en}{APSIPA Transactions
  on Signal and Information Processing}}, vol.~3, p. e14, 2014. [Online].
  Available:
  \url{https://www.cambridge.org/core/product/identifier/S2048770314000122/type/journal_article}
\BIBentrySTDinterwordspacing

\bibitem{lan_albert_2020}
\BIBentryALTinterwordspacing
Z.~Lan, M.~Chen, S.~Goodman, K.~Gimpel, P.~Sharma, and R.~Soricut,
  ``\BIBforeignlanguage{en}{{ALBERT}: {A} {Lite} {BERT} for {Self}-supervised
  {Learning} of {Language} {Representations}},''
  \emph{\BIBforeignlanguage{en}{arXiv:1909.11942 [cs]}}, Feb. 2020, arXiv:
  1909.11942. [Online]. Available: \url{http://arxiv.org/abs/1909.11942}
\BIBentrySTDinterwordspacing

\bibitem{ioffe2015batch}
S.~Ioffe and C.~Szegedy, ``Batch normalization: Accelerating deep network
  training by reducing internal covariate shift,'' \emph{arXiv preprint
  arXiv:1502.03167}, 2015.

\bibitem{elliott2020cyber}
D.~Elliott, E.~Martino, C.~E. Otero, A.~Smith, A.~M. Peter, B.~Luchterhand,
  E.~Lam, and S.~Leung, ``Cyber-physical analytics: Environmental sound
  classification at the edge,'' in \emph{2020 IEEE 6th World Forum on Internet
  of Things (WF-IoT)}.\hskip 1em plus 0.5em minus 0.4em\relax IEEE, 2020, pp.
  1--6.

\bibitem{oord2016wavenet}
A.~v.~d. Oord, S.~Dieleman, H.~Zen, K.~Simonyan, O.~Vinyals, A.~Graves,
  N.~Kalchbrenner, A.~Senior, and K.~Kavukcuoglu, ``Wavenet: A generative model
  for raw audio,'' \emph{arXiv preprint arXiv:1609.03499}, 2016.

\bibitem{sperber_self-attentional_2018}
\BIBentryALTinterwordspacing
M.~Sperber, J.~Niehues, G.~Neubig, S.~Stüker, and A.~Waibel,
  ``\BIBforeignlanguage{en}{Self-{Attentional} {Acoustic} {Models}},''
  \emph{\BIBforeignlanguage{en}{arXiv:1803.09519 [cs]}}, Jun. 2018, arXiv:
  1803.09519. [Online]. Available: \url{http://arxiv.org/abs/1803.09519}
\BIBentrySTDinterwordspacing

\bibitem{wu2016google}
Y.~Wu, M.~Schuster, Z.~Chen, Q.~V. Le, M.~Norouzi, W.~Macherey, M.~Krikun,
  Y.~Cao, Q.~Gao, K.~Macherey \emph{et~al.}, ``Google's neural machine
  translation system: Bridging the gap between human and machine translation,''
  \emph{arXiv preprint arXiv:1609.08144}, 2016.

\bibitem{nazabal2020handling}
A.~Nazabal, P.~M. Olmos, Z.~Ghahramani, and I.~Valera, ``Handling incomplete
  heterogeneous data using vaes,'' \emph{Pattern Recognition}, p. 107501, 2020.

\bibitem{zhao2020musicoder}
Y.~Zhao, X.~Wu, Y.~Ye, J.~Guo, and K.~Zhang, ``Musicoder: A universal
  music-acoustic encoder based on transformers,'' \emph{arXiv preprint
  arXiv:2008.00781}, 2020.

\bibitem{jiao2020translate}
Y.~Jiao, ``Translate reverberated speech to anechoic ones: Speech
  dereverberation with bert,'' \emph{arXiv preprint arXiv:2007.08052}, 2020.

\bibitem{lin2020mcunet}
J.~Lin, W.-M. Chen, Y.~Lin, J.~Cohn, C.~Gan, and S.~Han, ``Mcunet: Tiny deep
  learning on iot devices,'' \emph{arXiv preprint arXiv:2007.10319}, 2020.

\bibitem{coral}
\BIBentryALTinterwordspacing
``Retrain a classification model on-device with weight imprinting.'' [Online].
  Available:
  \url{https://coral.ai/docs/edgetpu/retrain-classification-ondevice/}
\BIBentrySTDinterwordspacing

\end{thebibliography}

\begin{IEEEbiography}[{\includegraphics[width=1in,height=1.25in,clip,keepaspectratio]{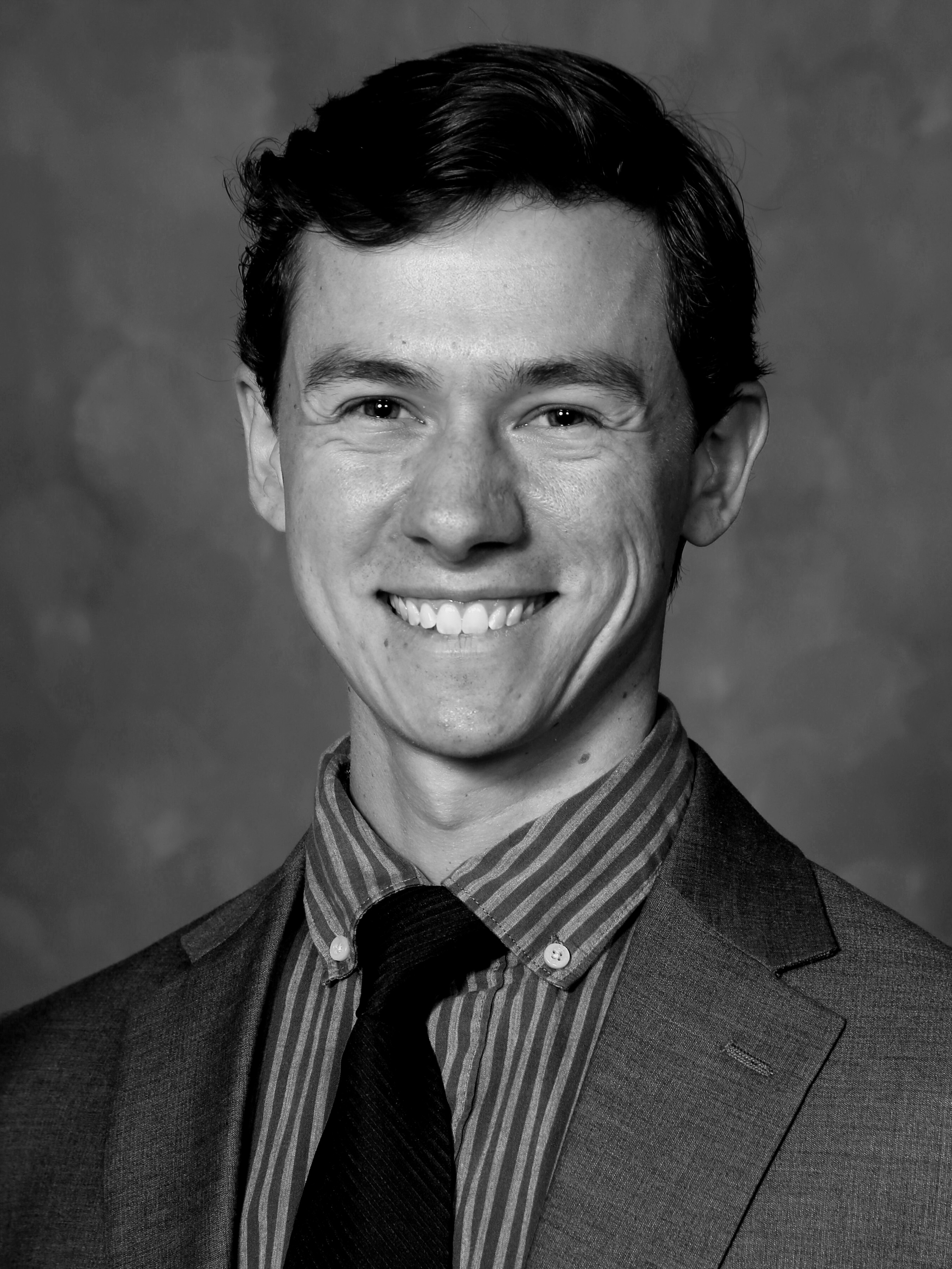}}]{David Elliott} (M’2017) is a Ph.D. candidate in Computer Engineering at Florida Institute of Technology. He completed his B.S. in Computer Engineering in 2017, and his M.S. in the same in 2018. He has three years of industry experience, where he has made contributions in the areas of cloud systems, cyber resiliency, machine learning, and the Internet of Things. His work has been presented to high-level government executives, and used to define and advance the state of the art in practice. He has authored several papers, appearing in IEEE CLOUD and IEEE World Forum on the Internet of Things.
\end{IEEEbiography}

\begin{IEEEbiography}[{\includegraphics[width=1in,height=1.25in,clip,keepaspectratio]{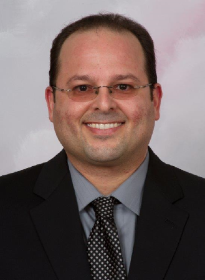}}]{Carlos E. Otero} (SM’09) received a B.S. degree in computer science, a M.S. degree in software engineering, a M.S. degree in systems engineering, and a Ph.D. degree in computer engineering from Florida Institute of Technology, Melbourne. 
He is currently Associate Professor and the Co-Director of the Center for Advanced Data Analytics and Systems (CADAS), Florida Institute of Technology. He was an Assistant Professor with the University of South Florida and the University of Virginia at Wise. He has authored over 70 papers in wireless sensor networks, Internet-of-Things, big data, and hardware/software systems. His research interests include performance analysis, evaluation, and optimization of computer systems, including wireless ad hoc and sensor networks. He has over twelve years of industry experience in satellite communications systems, command and control systems, wireless security systems, and unmanned aerial vehicle systems. 
\end{IEEEbiography}

\begin{IEEEbiography}[{\includegraphics[width=1in,height=1.25in,clip,keepaspectratio]{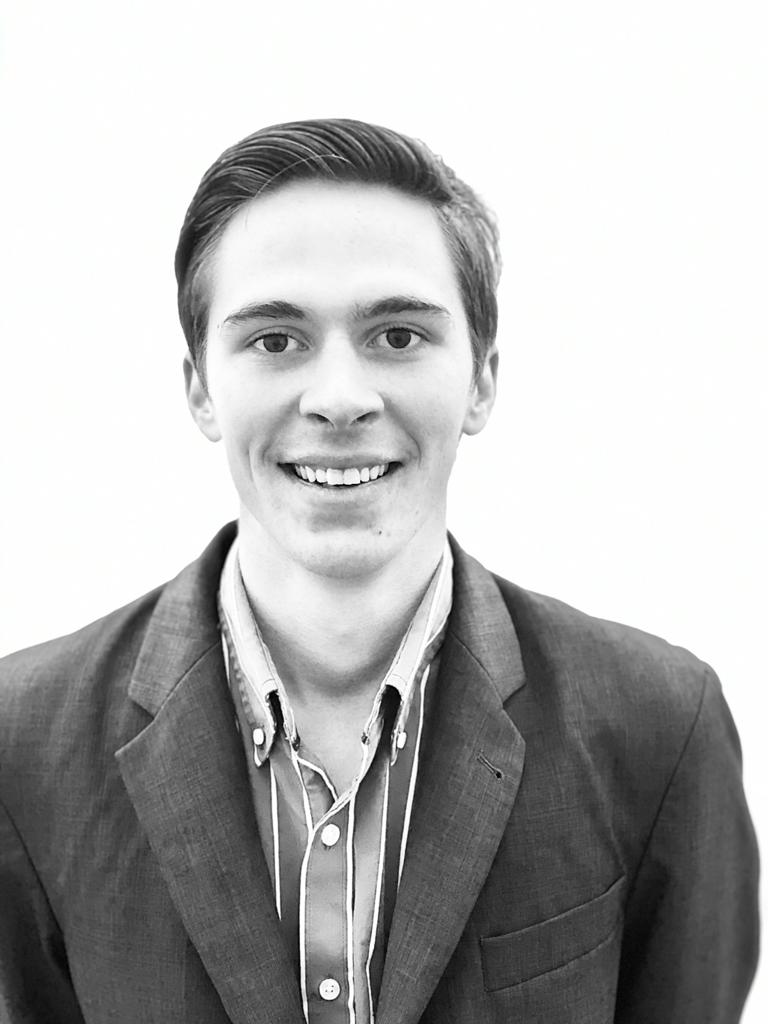}}]{Steven Wyatt} (M'2020) performs research and development at Northrop Grumman Corporation while completing his B.S. in Computer Engineering at Florida Institute of Technology. He has extensive experience in fuzzing, wireless systems, and machine learning. His work has been the basis for corporate AI strategy, and he continues to improve the state of the art in efficient edge analytics through his research.
\end{IEEEbiography}

\begin{IEEEbiography}[{\includegraphics[width=1in,height=1.25in,clip,keepaspectratio]{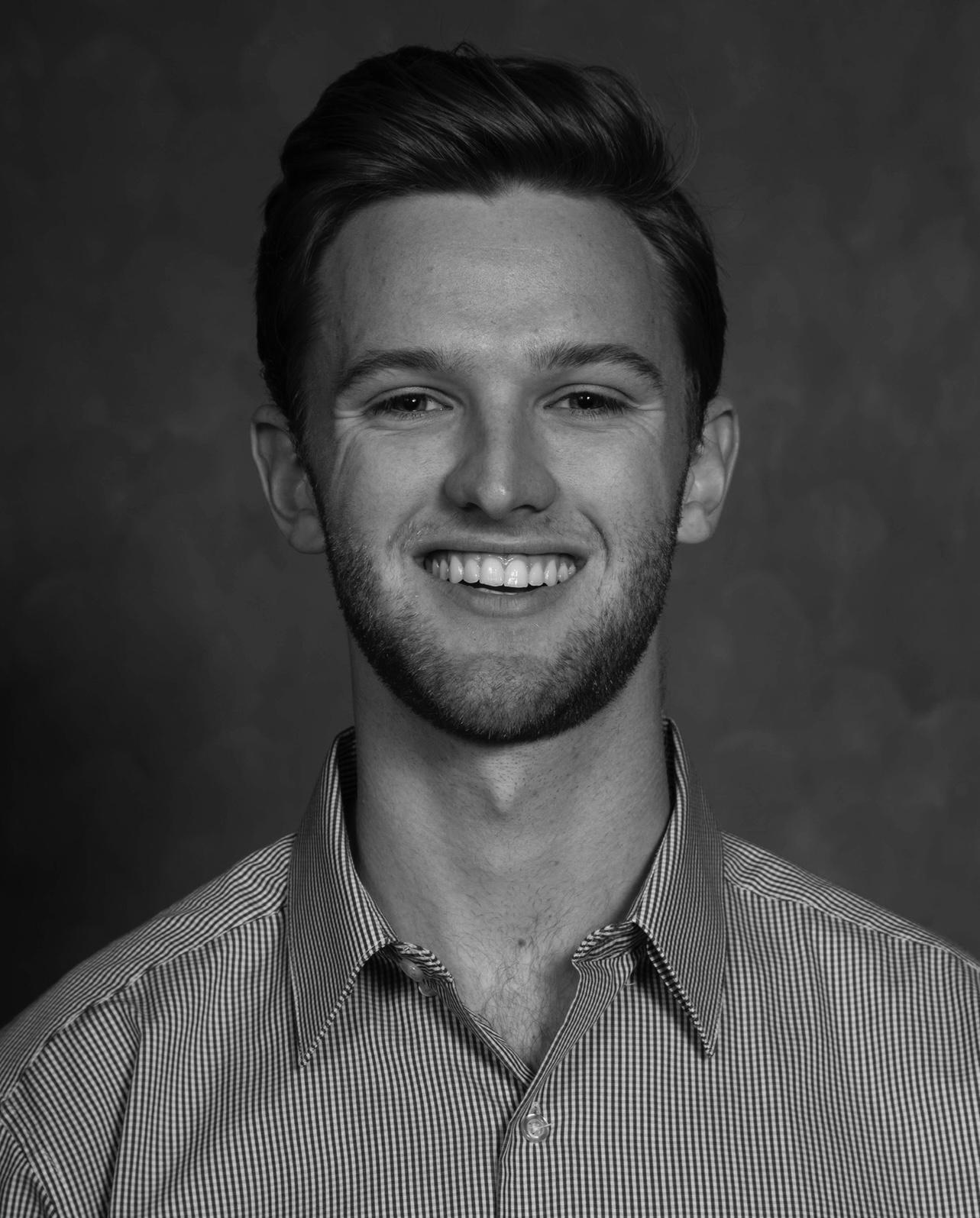}}]{Evan Martino} (M'2020) completed his B.S in Computer Engineering, and is pursuing his M.S in Computer Engineering at the Florida Institute of Technology. He has two years of industry experience, performing research and development in networking, distributed systems, and machine learning. He has been published in the World Forum on the Internet of things, and his work is used to provide state-of-the-art analytics in production systems today.
\end{IEEEbiography}

\end{document}